\documentclass{article}
\usepackage{spconf,amsmath,graphicx,bm}
\usepackage{cite,amssymb,amsfonts,textcomp,xcolor,booktabs}
\usepackage{url}

\title{MATE: Matryoshka Audio--Text Embeddings for Open-Vocabulary Keyword Spotting}
\name{Youngmoon Jung, Myunghun Jung, Joon-Young Yang, Yong-Hyeok Lee, Jaeyoung Roh, Hoon-Young Cho}
\address{AI Solution Team, Samsung Research, Seoul, South Korea}
\begin{document}
\ninept
\maketitle
\begin{abstract}
Open-vocabulary keyword spotting (KWS) with text-based enrollment has emerged as a flexible alternative to fixed-phrase triggers. 
Prior utterance-level matching methods, from an embedding-learning standpoint, learn embeddings at a single fixed dimensionality. 
We depart from this design and propose Matryoshka Audio--Text Embeddings (MATE), a dual-encoder framework that encodes multiple embedding granularities within a single vector via nested sub-embeddings (``prefixes''). Specifically, we introduce a PCA-guided prefix alignment: PCA-compressed versions of the full text embedding for each prefix size serve as teacher targets to align both audio and text prefixes. This alignment concentrates salient keyword cues in lower-dimensional prefixes, while higher dimensions add detail. MATE is trained with standard deep metric learning objectives for audio--text KWS, and is loss-agnostic. To our knowledge, this is the first application of matryoshka-style embeddings to KWS, achieving state-of-the-art results on WSJ and LibriPhrase without any inference overhead.
\end{abstract}
\section{Introduction}
\label{sec:intro}

Keyword spotting (KWS) enables devices to detect specific trigger words in speech, traditionally focusing on a small set of fixed keywords (e.g., ``Alexa'', ``OK Google'', or ``Hi Bixby'') \cite{Chen14-ICASSP, Sainath-INTERSPEECH, TANG17-ICASSP, Jung-ICASSP-TAadapter}. Recently, there is increasing interest in flexible, open-vocabulary KWS that allows users to define arbitrary keywords \cite{Chen-ICASSP, Huang-ICASSP, He-ICLR, Jung-INTERSPEECH-AsyP, Shin-INTERSPEECH, Nishu-INTERSPEECH, Lee-INTERSPEECH, RPL-Jung-INTERSPEECH, jin24d_interspeech, ADML-Jung-INTERSPEECH}. 
In particular, text-based enrollment, where a user types the desired trigger word or phrase instead of providing spoken examples, offers convenience and facilitates large-scale deployment, because new keywords can be added by encoding the typed text, eliminating the need for spoken enrollment samples \cite{He-ICLR, Jung-INTERSPEECH-AsyP, Shin-INTERSPEECH, Nishu-INTERSPEECH, Lee-INTERSPEECH, RPL-Jung-INTERSPEECH, jin24d_interspeech}. This audio--text scenario requires a shared representation between speech and text so that the system can decide whether an utterance contains the enrolled keyword by comparing their embeddings.

In text-enrolled open-vocabulary KWS, two matching paradigms are commonly discussed: (i) \emph{utterance-level matching}, which compares pooled audio and text embeddings for an entire word or short phrase, and (ii) \emph{phoneme-level matching}, which aligns acoustic and textual sequences at a finer temporal resolution using attention \cite{Shin-INTERSPEECH, Lee-INTERSPEECH}, dynamic programming \cite{Nishu-INTERSPEECH}, or CTC-style alignment \cite{jin24d_interspeech}. This work focuses on utterance-level matching because it requires no sequence-alignment module and supports open-set retrieval with a single pooled embedding per modality, an attractive property for low-latency, on-device KWS. Extending our approach to phoneme-level matching is left for future work.

Motivated by Matryoshka Representation Learning (MRL) \cite{kusupati2022mrl}, we propose \textit{Matryoshka Audio--Text Embeddings} (MATE), a dual-encoder framework that learns nested sub-embeddings (hereafter, \emph{prefixes}) formed by selecting the leading portion of a single embedding vector. 
By gradually extending this selected portion, the model represents multiple embedding granularities within a single vector. 
In particular, lower-dimensional prefixes capture condensed keyword information, while higher-dimensional prefixes add fine-grained detail. 
Training uses standard deep metric learning (DML) \cite{Wang-CVPR} objectives for audio--text KWS.
In our experiments, we adopt Relational Proxy Loss (RPL) \cite{RPL-Jung-INTERSPEECH} as the main loss.
Yet, MATE is loss-agnostic and consistently improves AP across diverse DML objectives (Sec.~\ref{subsec:wsj_main}).

Summarizing, this paper makes two contributions. 
First, we introduce MATE, to our knowledge the first application of matryoshka-style representation learning to KWS, enabling multiple embedding granularities within a single forward pass from one dual-encoder.
Second, we propose a PCA-guided prefix alignment that treats PCA-compressed text sub-embeddings as teachers and aligns both acoustic and text prefixes with a delayed schedule; this mitigates cross-scale interference observed with naive per-prefix supervision and yields consistent gains on WSJ and LibriPhrase under standard DML objectives, without changing inference cost.

\section{Related Work}
\label{sec:related_works}

\noindent\textbf{Text-enrolled utterance-level KWS.}
In utterance-level matching, an acoustic encoder and a text encoder are trained with DML so that audio/text embeddings from the \emph{same} keyword are pulled together while those from \emph{different} keywords are pushed apart, enabling open-set retrieval.
Early multi-view acoustic word embeddings relied on triplet losses \cite{He-ICLR}.
Proxy objectives remain effective, including Asymmetric Proxy loss (AsyP) \cite{Jung-INTERSPEECH-AsyP} and its adaptive margin-and-scale variant AdaMS \cite{Jung-INTERSPEECH-AdaMS}. 
RPL adds cross-modal structural alignment and reports strong gains on WSJ \cite{RPL-Jung-INTERSPEECH}. 
ADML combines modality-adversarial training with DML to reduce the cross-modal gap \cite{ADML-Jung-INTERSPEECH}. 
Under proxy-based training, the text embedding of each keyword acts as a class proxy/centroid that shapes the acoustic space.

\medskip
\noindent\textbf{Matryoshka-style embeddings.}
MRL \cite{kusupati2022mrl} encodes multiple granularities within a single vector by using nested sub-embeddings (``prefixes''), so low-dimensional prefixes capture compact cues while larger prefixes add detail, enabling multiple embedding dimensionalities without extra encoder passes. 
In speaker verification, M-Vec \cite{wang2024mvec} and DSM-GS \cite{park2025dsmegs} show that compact prefixes can maintain accuracy and benefit from dimension-specific training. 
ESE \cite{li2025ese} adopts a two-stage recipe, \emph{learn-to-express} and \emph{learn-to-compress}, to scale model depth and embedding size. 
Prior observations indicate that naively supervising all prefixes with the main loss can induce competition across prefix sizes, whereas compression-driven ordering of features tends to stabilize multi-embedding training \cite{li2025ese}. 
From an embedding-learning standpoint, prior utterance-level KWS systems typically learn a single fixed-dimensional embedding. We depart from this single-granularity paradigm by training nested sub-embeddings that capture multiple information granularities. At inference, however, we use only the full-dim embedding, which is identical to the approach in prior work.

\begin{figure}[t]
  \centering
  \includegraphics[width=\columnwidth]{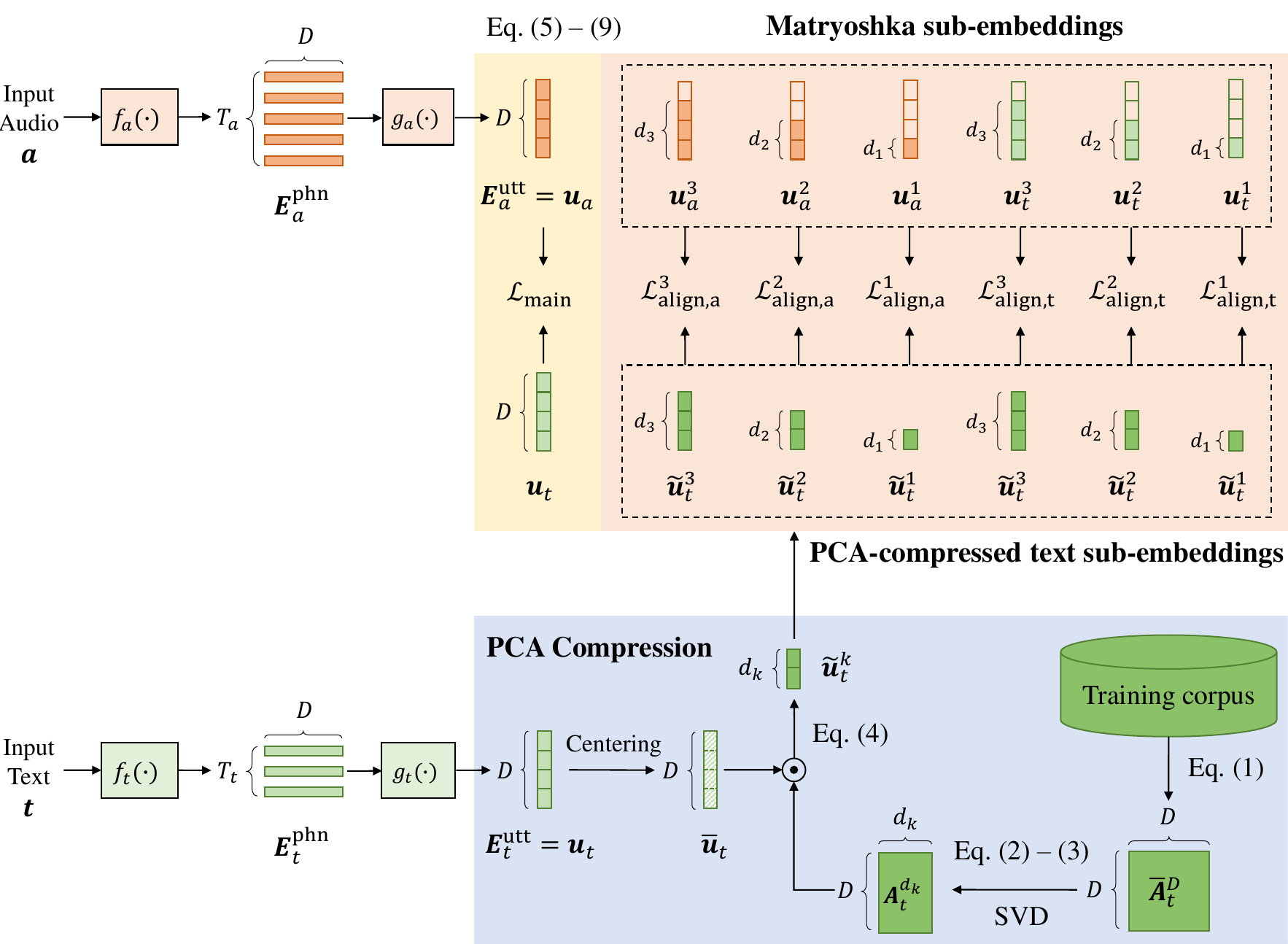}
  \vspace{-0.15cm}
    \caption{Overview of MATE. A corpus-wide text inner-dependency matrix $\bar{\boldsymbol{A}}^{D}_t$ is estimated and factorized (SVD) to obtain projection heads $\{\boldsymbol{A}^{d_k}_t\}$. For each $d_k$, the acoustic/text prefixes $\boldsymbol{u}^{k}_a,\boldsymbol{u}^{k}_t$ (leading $d_k$-dim sub-vectors) are aligned to the PCA-compressed text sub-embedding $\tilde{\boldsymbol{u}}^{k}_t$; the full-dim pair uses the main loss (RPL).}
  \label{fig:fig1}
  \vspace{-0.3cm}
\end{figure}

\section{Proposed Method}
\label{sec:method}

\subsection{Overview and Notation}

Fig.~\ref{fig:fig1} depicts the overall MATE pipeline.
Let $\mathcal{B}=\{(\boldsymbol{a}_i,\boldsymbol{t}_i,y_i)\}_{i=1}^N$ denote a mini-batch of audio, text, and keyword-label tuples.
For a sample $(\boldsymbol{a},\boldsymbol{t},y)\in\mathcal{B}$, the acoustic encoder $f_a(\cdot)$ and the text encoder $f_t(\cdot)$ produce phoneme-level embedding sequences
\[
\boldsymbol{E}^{\mathrm{phn}}_a=f_a(\boldsymbol{a}) \in \mathbb{R}^{T_a \times D},\qquad
\boldsymbol{E}^{\mathrm{phn}}_t=f_t(\boldsymbol{t}) \in \mathbb{R}^{T_t \times D},
\]
where $T_a$ and $T_t$ are sequence lengths and $D$ denotes the embedding dimensionality.
Following prior work \cite{RPL-Jung-INTERSPEECH,ADML-Jung-INTERSPEECH}, utterance-level embeddings are obtained via pooling:
\[
\boldsymbol{E}^{\mathrm{utt}}_a = g_a(\boldsymbol{E}^{\mathrm{phn}}_a) \in \mathbb{R}^{D},\qquad
\boldsymbol{E}^{\mathrm{utt}}_t = g_t(\boldsymbol{E}^{\mathrm{phn}}_t) \in \mathbb{R}^{D},
\]
where $g_a(\cdot)$ is channel- and context-dependent statistics pooling (CCSP) \cite{Desplanques-arxiv} and $g_t(\cdot)$ is global average pooling. Unless otherwise noted, all embeddings are $\ell_2$-normalized.

Nested sub-embeddings (i.e., \emph{prefixes}) are defined by taking the leading dimensions of a single vector.
Let $\mathcal{D}=\{d_1,\ldots,d_K\}$ be the set of prefix dimensionalities with $d_1<\cdots<d_K=D$.
Following the power-of-two (halving) schedule \cite{kusupati2022mrl},
\[
d_k \;=\; D \cdot 2^{-(K-k)}, \qquad k=1,\ldots,K,
\]
so that $\mathcal{D}=\{D/2^{K-1},\ldots,D/2,D\}$.
Given $\boldsymbol{u}_a \triangleq \boldsymbol{E}^{\mathrm{utt}}_a \in \mathbb{R}^{D}$ and
$\boldsymbol{u}_t \triangleq \boldsymbol{E}^{\mathrm{utt}}_t \in \mathbb{R}^{D}$,
the prefixes are
$\boldsymbol{u}_a^{k}=\boldsymbol{u}_a[1\!:\!d_k]$ and
$\boldsymbol{u}_t^{k}=\boldsymbol{u}_t[1\!:\!d_k]$.

We denote by $\mathcal{L}_{\mathrm{main}}$ the utterance-level audio--text objective applied to the full-dim pair $(\boldsymbol{u}_a,\boldsymbol{u}_t)$; it is instantiated later in Sec.~\ref{subsec:mate}.
In our experiments, we fix $D{=}256$ and vary the number of prefix sizes $K\in\{2,3,4,5\}$; for example, $K{=}5$ yields $\mathcal{D}=\{16,32,64,128,256\}$.

\subsection{Matryoshka Audio--Text Embeddings (MATE)}
\label{subsec:mate}

We motivate MATE with an \emph{information squeeze} perspective: the shared audio--text space should arrange keyword evidence so that lower-dimensional prefixes carry condensed, high-salience cues, while progressively larger prefixes add fine-grained detail. In proxy-based DML, text embeddings act as representative vectors for each keyword. MATE concentrates essential textual information into the leading dimensions of the acoustic embedding by aligning both acoustic and text \emph{prefixes} to \emph{PCA-compressed text sub-embeddings} derived from the full text embedding.

\smallskip
\noindent\textbf{Corpus-wide dependency and PCA compression.}
Let $\boldsymbol{\mu}^{D}_t\!\in\!\mathbb{R}^{D}$ be the mean of full text embeddings over the training corpus, and let $\bar{\boldsymbol{u}}_t=\boldsymbol{u}_t-\boldsymbol{\mu}^{D}_t$ denote the centered text embedding.
Motivated by the self-attention mechanism \cite{transformer}, we form an \emph{inner-dependency} matrix via a scaled dot-product with row-wise softmax. For numerical stability and efficiency, we estimate a \emph{training-corpus-wide} dependency at the beginning of each epoch $e$ using the current text encoder:
\begin{equation}\label{eq:depA_epoch}
\bar{\boldsymbol{A}}^{D}_t \;=\; \mathrm{Softmax}_{\mathrm{row}}\!\left(\frac{1}{M}\sum_{j=1}^{M}\frac{\bar{\boldsymbol{u}}^{(j)}_t\,\bar{\boldsymbol{u}}^{(j)\top}_t}{\sqrt{D}}\right)\! \in \mathbb{R}^{D\times D},
\end{equation}
where $\{\bar{\boldsymbol{u}}^{(j)}_t\}_{j=1}^{M}$ are centered full text embeddings over the \emph{training set}.
We then compute its \emph{singular value decomposition (SVD)}:
\begin{equation}\label{eq:svd}
\bar{\boldsymbol{A}}^{D}_t \;=\; \boldsymbol{U}\,\boldsymbol{\Sigma}\,\boldsymbol{V}^{\top},
\end{equation}
where $\boldsymbol{U},\boldsymbol{V}\!\in\!\mathbb{R}^{D\times D}$ are orthogonal and $\boldsymbol{\Sigma}\!\in\!\mathbb{R}^{D\times D}$ is diagonal with nonnegative singular values in descending order along the diagonal.
For a target prefix dimensionality $d_k$ ($1\!\le\! d_k\!\le\! D$), the \emph{top-$d_k$ principal dependency components} are defined as
\begin{equation}\label{eq:Ak}
\boldsymbol{A}^{d_k}_t \;=\; \boldsymbol{U}_{:,1:d_k}\,\boldsymbol{\Sigma}_{1:d_k,\,1:d_k} \;\in\; \mathbb{R}^{D\times d_k}.
\end{equation}
Applying $\boldsymbol{A}^{d_k}_t$ to the centered full text embedding yields the \emph{PCA-compressed text sub-embedding}:
\begin{equation}\label{eq:utk_comp}
\tilde{\boldsymbol{u}}^{k}_t \;=\; \left(\boldsymbol{A}^{d_k}_t\right)^{\top}\bar{\boldsymbol{u}}_t \;\in\; \mathbb{R}^{d_k}.
\end{equation}
For brevity, we refer to these SVD-based projections as ``PCA-compressed,'' since they play the role of principal, variance-ordered directions for prefix targets in practice.
Intuitively, $\tilde{\boldsymbol{u}}^{k}_t$ collects the $d_k$ most influential dependency directions of $\boldsymbol{u}_t$. 
To inject this squeezed information into the leading dimensions of $\boldsymbol{u}_a$ and $\boldsymbol{u}_t$, we define the alignment objectives below.

\begin{table*}[t]
\centering
\caption{WSJ AP (\%) under RPL with four strategies:
\textbf{RPL (full-only)}, no matryoshka prefixes;
\textbf{Per-prefix RPL (ours)}, apply the main loss (RPL) to each nested sub-embedding;
\textbf{Per-prefix RPL + PCA-guided alignment (ours)}, apply both per-prefix RPL and PCA-guided alignment on the same prefixes;
\textbf{MATE (ours)}, apply RPL only to the full embedding and PCA-guided alignment to prefixes.
$\Delta_{\text{full}}=\mathrm{AP}_{\text{method}}-\mathrm{AP}_{\text{full-only}}$,
$\Delta_{\text{per-prefix}}=\mathrm{AP}_{\text{method}}-\mathrm{AP}_{\text{Per-prefix RPL}}$ (percentage points, pp).}
\label{tab:abl_mrl}
\vspace{+0.2cm}
\setlength{\tabcolsep}{9pt}\renewcommand{\arraystretch}{1.0}
\footnotesize
\begin{tabular}{lcccccc}
\toprule
Method & Per-prefix RPL & PCA-guided alignment & AP (\%) & $\Delta_{\text{full}}$ & $\Delta_{\text{per-prefix}}$ \\
\midrule
RPL (full-only)                          & --         & --            & 78.66 & --     & --     \\
Per-prefix RPL (ours)                    & \checkmark & --            & 79.49 & +0.83  & --     \\
Per-prefix RPL + PCA-guided alignment (ours) & \checkmark & \checkmark    & 78.01 & -0.65  & -1.48  \\
\textbf{MATE (ours)}                     & --         & \checkmark    & \textbf{80.94} & \textbf{+2.28} & \textbf{+1.45} \\
\bottomrule
\end{tabular}
\vspace{-0.1cm}
\end{table*}

\smallskip
\noindent\textbf{Prefix alignment losses.}
For each prefix size \(d_k \in \mathcal{D}\setminus\{D\}\) (i.e., for all nested sub-embeddings except the full dim \(D\)), we align the PCA-compressed text sub-embedding \(\tilde{\boldsymbol{u}}^{k}_t\) (teacher) with both the acoustic prefix \(\boldsymbol{u}_a^{k}\) and the text prefix \(\boldsymbol{u}_t^{k}\) (students).
This encourages (i) the leading acoustic coordinates to absorb the compressed, high-salience keyword cues, and (ii) the compressed text proxy to remain consistent with the geometry of the original text prefix at the same dimensionality.
Let \(\phi_{\tau}(\boldsymbol{x})=\mathrm{Softmax}(\boldsymbol{x}/\tau)\) be a temperature-softened softmax.
Each alignment combines an MSE term and a KL divergence between softened distributions, motivated by \cite{kim-IJCAI}:
\begin{align}
\label{eq:align_a_k}
\mathcal{L}_{\mathrm{align},a}^{k}
&= \mathrm{MSE}\!\left(\boldsymbol{u}_{a}^{k},\,\tilde{\boldsymbol{u}}_{t}^{k}\right)
+ \mathrm{KL}\!\left(\phi_{\tau}(\boldsymbol{u}_{a}^{k}) \,\big\|\, \phi_{\tau}(\tilde{\boldsymbol{u}}_{t}^{k})\right),\\
\label{eq:align_t_k}
\mathcal{L}_{\mathrm{align},t}^{k}
&= \mathrm{MSE}\!\left(\boldsymbol{u}_{t}^{k},\,\tilde{\boldsymbol{u}}_{t}^{k}\right)
+ \mathrm{KL}\!\left(\phi_{\tau}(\boldsymbol{u}_{t}^{k}) \,\big\|\, \phi_{\tau}(\tilde{\boldsymbol{u}}_{t}^{k})\right),\\
\label{eq:align_k}
\mathcal{L}_{\mathrm{align}}^{k}
&= \mathcal{L}_{\mathrm{align},a}^{k} + \mathcal{L}_{\mathrm{align},t}^{k}.
\end{align}
Define \(\mathcal{K}=\{1,\dots,K-1\}\).
The total alignment is then
\begin{equation}\label{eq:align_total}
\mathcal{L}_{\mathrm{align}}=\sum_{k\in\mathcal{K}}\mathcal{L}_{\mathrm{align}}^{k}.
\end{equation}

\smallskip
\noindent\textbf{Final objective and schedule.}
Let $\mathcal{L}_{\mathrm{main}}$ be the utterance-level audio--text objective on the full pair $(\boldsymbol{u}_a,\boldsymbol{u}_t)$ (instantiated as RPL in our experiments). The final loss is
\begin{equation}\label{eq:total_obj}
\mathcal{L}_{\mathrm{total}} \;=\; \mathcal{L}_{\mathrm{main}} \;+\; \lambda_{\mathrm{align}}\,\mathcal{L}_{\mathrm{align}}.
\end{equation}
We use a delayed schedule
\begin{equation}\label{eq:lambda_schedule}
\lambda_{\mathrm{align}}(e)=
\begin{cases}
0, & 1 \le e \le 20,\\[2pt]
0.5, & e \ge 21,
\end{cases}
\end{equation}
where $e$ is the epoch index. This lets the full text embedding first become a reliable proxy/centroid under $\mathcal{L}_{\mathrm{main}}$ before the compression-based alignment starts steering the prefixes.

\smallskip
\noindent\textbf{Epoch routine.}
At the \emph{start} of each epoch, we recompute the corpus statistics with the current text encoder $f_t$: (i) the corpus mean $\boldsymbol{\mu}^{D}_t$, (ii) the corpus-wide dependency $\bar{\boldsymbol{A}}^{D}_t$ via Eq.~\eqref{eq:depA_epoch}, and (iii) the projection heads $\{\boldsymbol{A}^{d_k}_t\}_{k=1}^{K-1}$ by SVD as in Eqs.~\eqref{eq:svd}--\eqref{eq:Ak}.
During training, for each mini-batch we obtain PCA-compressed targets \emph{on the fly} by applying Eq.~\eqref{eq:utk_comp} to the batch's centered text embeddings $\bar{\boldsymbol{u}}_t$, and then compute the prefix alignment losses in Eqs.~\eqref{eq:align_a_k}--\eqref{eq:align_k}. The alignment term is weighted according to the schedule in Eq.~\eqref{eq:lambda_schedule}.

\section{Experiments}

\subsection{Experimental Setup}

\noindent\textbf{Datasets and evaluation protocol.}
We train on King-ASR-066 \cite{Speechocean-DB} following the setup of \cite{RPL-Jung-INTERSPEECH}, including the same augmentation recipe. This yields roughly 4.6k hours of word-level segments covering about 210k word classes. 
For evaluation, we use Wall Street Journal (WSJ) \cite{Paul-WSN} and LibriPhrase \cite{Shin-INTERSPEECH}. WSJ trials are constructed as in \cite{RPL-Jung-INTERSPEECH}: 3.2k distinct words and 18k word segments, further corrupted with RIRs \cite{Ko-ICASSP} and MUSAN noises \cite{Snyder-arxiv}. The WSJ test comprises 18k positive pairs and 50$\times$ more negatives to emulate real-world imbalance; we therefore report Average Precision (AP). 
LibriPhrase follows its official splits and metrics (Equal Error Rate, EER; Area Under the ROC Curve, AUC) on LP\textsubscript{E} and LP\textsubscript{H}.

\smallskip
\noindent\textbf{Inputs and encoders.}
Acoustic features are 40-d log-Mel filterbanks with 25\,ms windows, 10\,ms frame shift, and utterance-level mean normalization. 
The acoustic encoder is a 256-channel ECAPA-TDNN \cite{Desplanques-arxiv} ($\sim$1.8M parameters), a standard choice in recent KWS works \cite{Li-INTERSPEECH, RPL-Jung-INTERSPEECH, Jung-ICASSP-TAadapter, ADML-Jung-INTERSPEECH}.
Text is tokenized with a grapheme-to-phoneme (G2P) front end \cite{ParkG2P} and embedded via a trainable 256-d lookup table; a 2-layer bi-LSTM (256 hidden units), followed by global average pooling and a linear layer, produces a 256-d text embedding.

\smallskip
\noindent\textbf{Optimization, batching, and inference.}
All models are implemented in PyTorch and optimized with AdamW (initial learning rate $10^{-4}$; weight decay $10^{-5}$) for 100 epochs. Each mini-batch contains 500 utterances from 250 keywords (two utterances per keyword). The main utterance-level objective is RPL with the same hyperparameters as \cite{RPL-Jung-INTERSPEECH}. For the MATE alignment, we use temperature $\tau{=}1$ in Eqs.~(\ref{eq:align_a_k})-(\ref{eq:align_k}) and apply the loss-weight schedule for $\lambda_{\mathrm{align}}$ as in Sec.~\ref{subsec:mate}. Inference computes cosine similarity between the full-dim utterance-level embeddings $\boldsymbol{u}_a$ and $\boldsymbol{u}_t$.
Training and evaluation run on two H100 GPUs (about one day for 100 epochs). By default we set $D{=}256$ and specify the matryoshka prefix set $\mathcal{D}$ (i.e., $K$ and $\{d_k\}$) per experiment. For baseline objectives (e.g., Proxy\text{-}BD (Binomial Deviance) \cite{Yi-ICPR}, Proxy\text{-}MS (Multi\text{-}Similarity) \cite{Wang-CVPR}, CLAT \cite{Kewei-arxiv}, AsyP \cite{Jung-INTERSPEECH-AsyP}, AdaMS \cite{Jung-INTERSPEECH-AdaMS}, and RPL \cite{RPL-Jung-INTERSPEECH}), unless otherwise stated we follow the settings in their original papers.

\subsection{Multi-scale supervision under RPL}
We first analyze different multi-scale supervision schemes under the RPL main loss. Table~\ref{tab:abl_mrl} compares four strategies: (i) a baseline \textit{RPL (full-only)} model without nested sub-embeddings; (ii) \textit{Per-prefix RPL}, which applies the main loss to each prefix; (iii) a combination of per-prefix RPL and our PCA-guided alignment; and (iv) \textit{MATE}, which reserves RPL for the full embedding and uses PCA-guided alignment only on prefixes.

Across these variants, two of our multi-scale approaches, Per-prefix RPL and MATE, outperform the full-only baseline, indicating the benefit of leveraging nested sub-embeddings during training.
Per-prefix RPL yields a modest gain over full-only (+0.83\,pp), while MATE provides the largest improvement (+2.28\,pp) and surpasses Per-prefix RPL by +1.45\,pp. 
In contrast, combining Per-prefix RPL with PCA-guided alignment underperforms even the full-only baseline ($-0.65$\,pp), suggesting conflicting supervision when both objectives are imposed on the same prefixes. 
We hypothesize that Per-prefix RPL pushes every prefix to be fully discriminative, whereas PCA-guided alignment pulls them toward compressed proxies emphasizing variance ordering, increasing interference and optimization difficulty.
By reserving the main loss for the full vector and using compression-guided targets as prefix regularizers, MATE avoids these conflicts and achieves the best AP. 

\begin{table}[t]
\centering
\caption{WSJ test AP (\%). Baselines vs.\ MATE. MATE uses $K{=}5$ with $\mathcal{D}=\{16,32,64,128,256\}$. 
$\Delta$ denotes absolute change in AP (percentage points): $\Delta=\mathrm{AP}_{\text{MATE}}-\mathrm{AP}_{\text{baseline}}$.}
\vspace{+0.1cm}
\label{tab:main_wsj}
\setlength{\tabcolsep}{5pt}\renewcommand{\arraystretch}{1.0}
\footnotesize
\begin{tabular}{lccc}
\toprule
Objective (utterance-level) & Baseline AP & MATE AP & $\Delta$ \\
\midrule
Proxy-BD \cite{Yi-ICPR}            & 69.76 & 72.95 & {+}3.19 \\
Proxy-MS \cite{Wang-CVPR}          & 70.18 & 73.12 & {+}2.94 \\
CLAT \cite{Kewei-arxiv}                   & 70.52 & 73.53 & {+}3.01 \\
AsyP \cite{Jung-INTERSPEECH-AsyP}         & 71.66 & 73.98 & {+}2.32 \\
AdaMS \cite{Jung-INTERSPEECH-AdaMS}       & 73.87 & 75.68 & {+}1.81 \\
RPL \cite{RPL-Jung-INTERSPEECH}           & 78.66 & \textbf{80.94} & \textbf{+}2.28 \\
\bottomrule
\end{tabular}
\vspace{-0.2cm}
\end{table}

\subsection{WSJ main results}\label{subsec:wsj_main}
Table~\ref{tab:main_wsj} reports WSJ AP (\%) for six utterance-level objectives---Proxy-BD and Proxy-MS (reformulations of BD and MS \cite{Jung-INTERSPEECH-AsyP}), CLAT, AsyP, AdaMS, and RPL---and their MATE-enhanced counterparts.
All systems share the same data, encoders, and batching; only the utterance-level objective ($\mathcal{L}_{\mathrm{main}}$) differs. 
We set $K{=}5$ with $\{16,32,64,128,256\}$ for the main comparison, following common practice in prior work \cite{park2025dsmegs}, before ablating this choice in Sec.~\ref{subsec:num_prefixes}.

MATE improves AP across all objectives (avg +2.59\,pp; +1.81--+3.19), with the largest gain on Proxy\text{-}BD (+3.19\,pp). Even the strongest baseline, RPL, rises to 80.94\% AP (+2.28\,pp), confirming that nested sub-embeddings complement diverse metric-learning objectives (relative +2.4--4.6\%).

\subsection{Effect of the number of prefixes $K$}\label{subsec:num_prefixes}
For the main WSJ comparison, we set $K{=}5$ following prior practice in matryoshka-style encoders (e.g., DSM-GS \cite{park2025dsmegs}), and we additionally ablate $K$ while keeping $D{=}256$ and the RPL main loss fixed.
Table~\ref{tab:num_scales} reports WSJ AP. 
All multi-scale \textbf{MATE} settings outperform the \textit{RPL (full-only)} baseline.

The best AP is obtained with $K{=}3$ ($\{64,128,256\}$). 
Moving to $K{=}4$ reduces AP, consistent with the intuition that very small dimensions (32-d) can limit discriminative capacity, a tendency also noted in a different speech task \cite{park2025dsmegs}. 
Adding a 16-d prefix ($K{=}5$) raises AP relative to $K{=}4$; we conjecture that the extra coarse prefix distributes alignment pressure across the prefixes and alleviates over-regularization at 32-d. 
This demonstrates the general robustness of the multi-prefix approach, as the performance at $K{=}5$ remains highly competitive and only slightly below the peak at $K{=}3$ (80.94\% vs.\ 81.03\%).
In further experiments, we observed no additional AP gains for $K\!\ge\!6$.


\begin{table}[t]
\centering
\caption{WSJ AP (\%) for MATE (ours) with RPL as the main loss under varying numbers of prefixes $K$ (same $\lambda_{\mathrm{align}}$ schedule). 
$\Delta_{\text{full}}=\mathrm{AP}-\mathrm{AP}_{\text{RPL(full-only)}}$ in percentage points (pp).}
\label{tab:num_scales}
\vspace{+0.1cm}
\setlength{\tabcolsep}{5.5pt}\renewcommand{\arraystretch}{1.15}
\footnotesize
\begin{tabular}{lcc}
\toprule
Configuration & AP (\%) & $\Delta_{\text{full}}$ (pp) \\
\midrule
\textit{RPL (full-only)}, $K{=}1:\{256\}$ (no prefixes) & 78.66 & -- \\
\textbf{MATE (ours)}, $K{=}2:\{128,256\}$ & 79.02 & +0.36 \\
\textbf{MATE (ours)}, $K{=}3:\{64,128,256\}$ & \textbf{81.03} & \textbf{+2.37} \\
\textbf{MATE (ours)}, $K{=}4:\{32,64,128,256\}$ & 79.38 & +0.72 \\
\textbf{MATE (ours)}, $K{=}5:\{16,32,64,128,256\}$ & 80.94 & +2.28 \\
\bottomrule
\end{tabular}
\vspace{-0.1cm}
\end{table}

\subsection{Effect of alignment composition}
We ablate alignment composition by varying the weights between MSE and temperature-softened KL in Eqs.~(\ref{eq:align_a_k})--(\ref{eq:align_k}).
Unless noted, all runs use the same $\lambda_{\mathrm{align}}$ schedule and $K{=}3$ prefixes $\{64,128,256\}$ with temperature $\tau{=}1$. 
Table~\ref{tab:align_components} shows that the equal combination (1:1) yields the best AP on WSJ.

\begin{table}[t]
\centering
\caption{WSJ AP (\%) for MATE (ours) under different alignment compositions (MSE:KL) with RPL main loss, $K{=}3$ prefixes $\{64,128,256\}$, and the same $\lambda_{\mathrm{align}}$ schedule. 
$\Delta_{\text{MSE}}=\mathrm{AP}-\mathrm{AP}_{\text{MSE-only}}$, 
$\Delta_{\text{KL}}=\mathrm{AP}-\mathrm{AP}_{\text{KL-only}}$ in percentage points (pp).}
\label{tab:align_components}
\vspace{+0.1cm}
\setlength{\tabcolsep}{6pt}\renewcommand{\arraystretch}{1.0}
\footnotesize
\begin{tabular}{lccc}
\toprule
MSE:KL & AP (\%) & $\Delta_{\text{MSE}}$ (pp) & $\Delta_{\text{KL}}$ (pp) \\
\midrule
1:0 (MSE only)  & 79.96 & --    & --    \\
0:1 (KL only)   & 80.04 & +0.08 & --    \\
1:1 (proposed)  & \textbf{81.03} & \textbf{+1.07} & \textbf{+0.99} \\
\bottomrule
\end{tabular}
\vspace{-0.2cm}
\end{table}

The two terms provide complementary signals: MSE enforces proximity in Euclidean space, while the softened KL divergence aligns the relative importance of each dimension. Combining them yields more robust improvements than using either term alone, which supports our goal of using PCA-guided targets to concentrate salient information into the prefixes.

\subsection{Performance on LibriPhrase}
Table~\ref{tab:tab4} compares EER (\%) and AUC (\%) on LibriPhrase's easy (LP\textsubscript{E}) and hard (LP\textsubscript{H}) splits.
To assess cross-corpus generalization, all systems are trained on King-ASR-066 and evaluated on LibriPhrase without using its training set.
For a fair comparison, all systems utilize the same acoustic and text encoders as MATE.
Furthermore, to isolate the effect of the DML objective, we re-implement ADML \cite{ADML-Jung-INTERSPEECH} \emph{without} the SphereFace2-based auxiliary loss (SF2) used in its original implementation. MATE adopts RPL as its main loss (K=3), consistent with the best-performing WSJ model. Lower EER and higher AUC are better.

Relative to ADML (w/o SF2), MATE reduces EER by $-0.05$\,pp on LP\textsubscript{E} (1.38 vs.\ 1.43) and by $-0.98$\,pp on LP\textsubscript{H} (20.06 vs.\ 21.04), and increases AUC by $+0.02$\,pp on LP\textsubscript{E} (99.86 vs.\ 99.84) and $+1.79$\,pp on LP\textsubscript{H} (88.70 vs.\ 86.91). 
These cross-corpus gains mirror WSJ trends, indicating that matryoshka-style nested sub-embeddings with PCA-guided alignment improve generalization in text-enrolled KWS.

\begin{table}[t]
\centering
\caption{LibriPhrase results. EER (\%) and AUC (\%) on LP\textsubscript{E} (easy) and LP\textsubscript{H} (hard). 
All systems are trained on King-ASR-066.}
\label{tab:tab4}
\vspace{+0.1cm}
\setlength{\tabcolsep}{6pt}\renewcommand{\arraystretch}{1.1}
\footnotesize
\begin{tabular}{l|cc|cc}
\hline
Method & \multicolumn{2}{c|}{EER (\%)} & \multicolumn{2}{c}{AUC (\%)} \\
\cline{2-5}
 & LP\textsubscript{E} & LP\textsubscript{H} & LP\textsubscript{E} & LP\textsubscript{H} \\
\hline
CMCD \cite{Shin-INTERSPEECH}        & 4.20 & 25.94 & 99.12 & 81.14 \\
PhonMatchNet \cite{Lee-INTERSPEECH} & 2.33 & 24.11 & 99.70 & 83.39 \\
RPL \cite{RPL-Jung-INTERSPEECH}     & 1.54 & 22.45 & 99.84 & 85.31 \\
ADML (w/o SF2) \cite{ADML-Jung-INTERSPEECH} & 1.43 & 21.04 & 99.84 & 86.91 \\
\hline\hline
\textbf{MATE (ours)}                 & \textbf{1.38} & \textbf{20.06} & \textbf{99.86} & \textbf{88.70} \\
\hline
\end{tabular}
\vspace{-0.1cm}
\end{table}

\section{Conclusion}

We presented Matryoshka Audio--Text Embeddings (MATE), a dual-encoder framework that, to our knowledge, applies matryoshka-style representations to KWS for the first time. MATE organizes keyword information across nested sub-embeddings by placing the main metric-learning loss only on the full embedding and aligning both acoustic and text prefixes to PCA-compressed text sub-embeddings with a delayed weight schedule. 
Across benchmarks (WSJ and LibriPhrase), MATE improved AP on WSJ and EER/AUC on LibriPhrase over prior utterance-level baselines.
Ablations show that supervising prefixes alone already outperforms a full-only baseline, highlighting the value of nested sub-embeddings; using PCA-guided targets without per-prefix main losses stabilizes multi-scale training and yields the best results. 
In future work, we will extend MATE to phoneme-level matching and explore replacing PCA with more discriminative compression such as LDA.

\clearpage
\bibliographystyle{IEEEbib}
\bibliography{strings,refs}

\end{document}